\documentclass[10pt, conference]{IEEEtran}
\IEEEoverridecommandlockouts
\usepackage{cite}
\usepackage[colorlinks=true, urlcolor=blue, linkcolor=red]{hyperref}
\usepackage{amsmath,amssymb,amsfonts}
\usepackage{algorithmic}
\usepackage{graphicx}
\usepackage{textcomp}
\usepackage{xcolor}
\usepackage[most]{tcolorbox}

\def\BibTeX{{\rm B\kern-.05em{\sc i\kern-.025em b}\kern-.08em
    T\kern-.1667em\lower.7ex\hbox{E}\kern-.125emX}}

\newtcolorbox{findingbox}{
    colback=gray!10,           % Light gray background
    colframe=gray!50,          % Gray border
    boxrule=1pt,               % Border thickness
    arc=4mm,                   % Rounded corners
    left=6pt,                  % Left padding
    right=6pt,                 % Right padding
    top=6pt,                   % Top padding
    bottom=6pt,                % Bottom padding
    boxsep=0pt,                % Additional separation
    enhanced,                  % Enable advanced features
    breakable                  % Allow box to break across pages
}

\begin{document}

\title{TrustedCode: Evaluating Large Language Models for Developing Trusted Code through Automated Program Verification
}

\author{\IEEEauthorblockN{Muhammad Danial}
\IEEEauthorblockA{\textit{School of Computing and Mathematical Sciences} \\
\textit{University of Greenwich}\\
London, UK \\
md7114u@gre.ac.uk}
\and
\IEEEauthorblockN{Muhammad Taimoor Khan}
\IEEEauthorblockA{\textit{Centre for Sustainable Cyber Security} \\
\textit{University of Greenwich}\\
London, UK \\
m.khan@greenwich.ac.uk}
}

\maketitle

\begin{abstract}
% "Who is using AI to code?" reports that approx. 30\% of the github Python code contributed by only US users has been generated by AI, which is inherently unreliable, giving way to formal verification as a method to establish trust in the code. 
% The use of LLM-driven formal verification to ensure trusted software is a widely researched topic, but due to the complexity of writing specification and verifying that the software meets its specification, it is underutilized in practical applications. Furthermore, current tools only support the verification of Java and C programs. Therefore, we present TrustedCode to establish trust in AI-generated Python code through automated program verification. We harness LLMs to (i) type-check an AI-generated Python program, (ii) translate the well-typed program into WhyML specified program using an intermediate verification framework Why3, and (iii) verify that the translated program meets the specification using back-end automated theorem provers of Why3. Specifically, we provide a chain of thought approach using a LangGraph framework to identify the capability gaps of LLMs in writing WhyML specification. This will assist software developers in improving prompting and fine-tuning, resulting in generating and consequently using a reliable AI-generated code. To assess capability gaps, we tested 10 Python programs with varying complexity. The results showed that 50\% of all errors occurred due to domain gaps, roughly 30\% due to syntactic gaps and 20\% due to semantic and typing gaps. 
Recent reports, such as "Who is using AI to code?", indicate that approximately 30\% of Python code contributions to GitHub are AI-generated. Given the inherent unreliability of this code, formal verification emerges as a promising approach to establish trust in software correctness. Although leveraging large language models (LLMs) for formal verification has garnered increasing attention, practical adoption remains limited due to the challenges in writing formal specifications and verifying program correctness against them. Existing tools predominantly support verification for languages such as Java and C, with minimal support for Python. We present TrustedCode, an automated framework for verifying AI-generated Python code. TrustedCode uses LLMs to (i) type-check Python code, (ii) translate it into WhyML via the Why3 framework, and (iii) verify it against formal specifications using Why3 back-end automated theorem provers. Our method integrates chain-of-thought prompting through LangGraph to identify and address LLM limitations in specification generation and proving the specified programs. To evaluate TrustedCode, we conducted experiments on ten Python programs of varying complexity. Our analysis reveals that 50\% of the observed errors stem from domain knowledge gaps, 30\% from syntactic inaccuracies, and the remaining 20\% from semantic and type-related issues. These findings underscore the importance of targeted improvements in LLM-guided formal verification workflows for Python.
TrustedCode is an open source tool available at: \href{github.com/mtaimoorkhan/TrustedCode}{github.com/mtaimoorkhan/TrustedCode}. The video demonstration is available at: \href{youtu.be/Nr5I2g8ec-s}{youtu.be/Nr5I2g8ec-s}

\end{abstract}

% \begin{IEEEkeywords}
% Why3, LLM, Verification, LangGraph
% \end{IEEEkeywords}

\section{Introduction}
Formal reasoning about software is a widely used technique to establish trust in the software. Owing to the difficulties in writing formal specification and establishing corresponding proofs, it has not been widely accepted in industry~\cite{b1}. Since software complexity is increasing, traditional testing is becoming increasingly inadequate, necessitating formal verification to improve software reliability.

Human and complier feedback to repair proofs has demonstrated efficacy in several studies. For instance, \cite{b2} exhibited automatic proof repair using Coq to maintain verified systems. In \cite{b3}, the premise selection task is devolved to automated theorem proving while the proof generation strategy is handled by the LLM’s. While \cite{b4} introduces Baldur, which generates complete proofs using LLM’s fine-tuned on formal verification, and attempts to correct the proof in an iterative manner. 

In this context, there is potential for LLMs to be used not only for code generation, but also for error correction and ensuring that the relevant predicates and theories are developed. Therefore, in this paper, we aim to answer a research question as follows:

What are the capability gaps of a state-of-the-art LLM (Anthropic Claude Sonnet 4) in translating python programs into a automatically verifiable WhyML specified code, and what is its effectiveness when operating within a self-correction loop?

To answer this question, we present TrustedCode, a framework within the confines of a closed-loop system to translate Python program into WhyML specification language, noted for its strengths as a formal verification framework and support for back-end provers. This will incorporate an LLM, the Why3 software as well as an iterative repair sequence using LangGraph to produce a provably correct program. To correct LLM errors, we integrate a method to identify and classify capability gaps. This method moves beyond the current binary pass-fail metric to a concrete and structured classification methodology to specifically isolate the LLMs gaps in its understanding of types, theory, syntax, and semantics. Finally, we provide a practical analysis on the performance of the framework, including evaluating the capacity gaps of the LLM and the way they were fixed. This provided a holistic overview of the current state and limitations of LLM models for this application.
% \end{enumerate}

% 1. We propose a framework within the confines of a closed-loop system to translate Python into WhyML specification language, noted for its strengths as a formal verification framework and support for back-end provers. This will incorporate an LLM, the Why3 software as well as an iterative repair sequence using LangGraph to produce a provably correct program.

% 2. The LLM error correction process will integrate a method to identify and classify capability gaps. This method moves beyond the current binary pass-fail metric to a concrete and structured classification methodology to specifically isolate the LLMs gaps in its understanding of theory, syntax, and semantics. 

% 3. Finally, we provide a detailed practical analysis on the performance of the framework, including evaluating the capacity gaps of the LLM and the way they were fixed, whether by the LLM itself or by manual intervention (Human in the Loop). This provided a holistic overview of the current state and limitations of LLM models for this application.

\section{Methodology}

\subsection{System Architecture}

\begin{figure*}[t]
    \centering
    \includegraphics[scale=0.36]{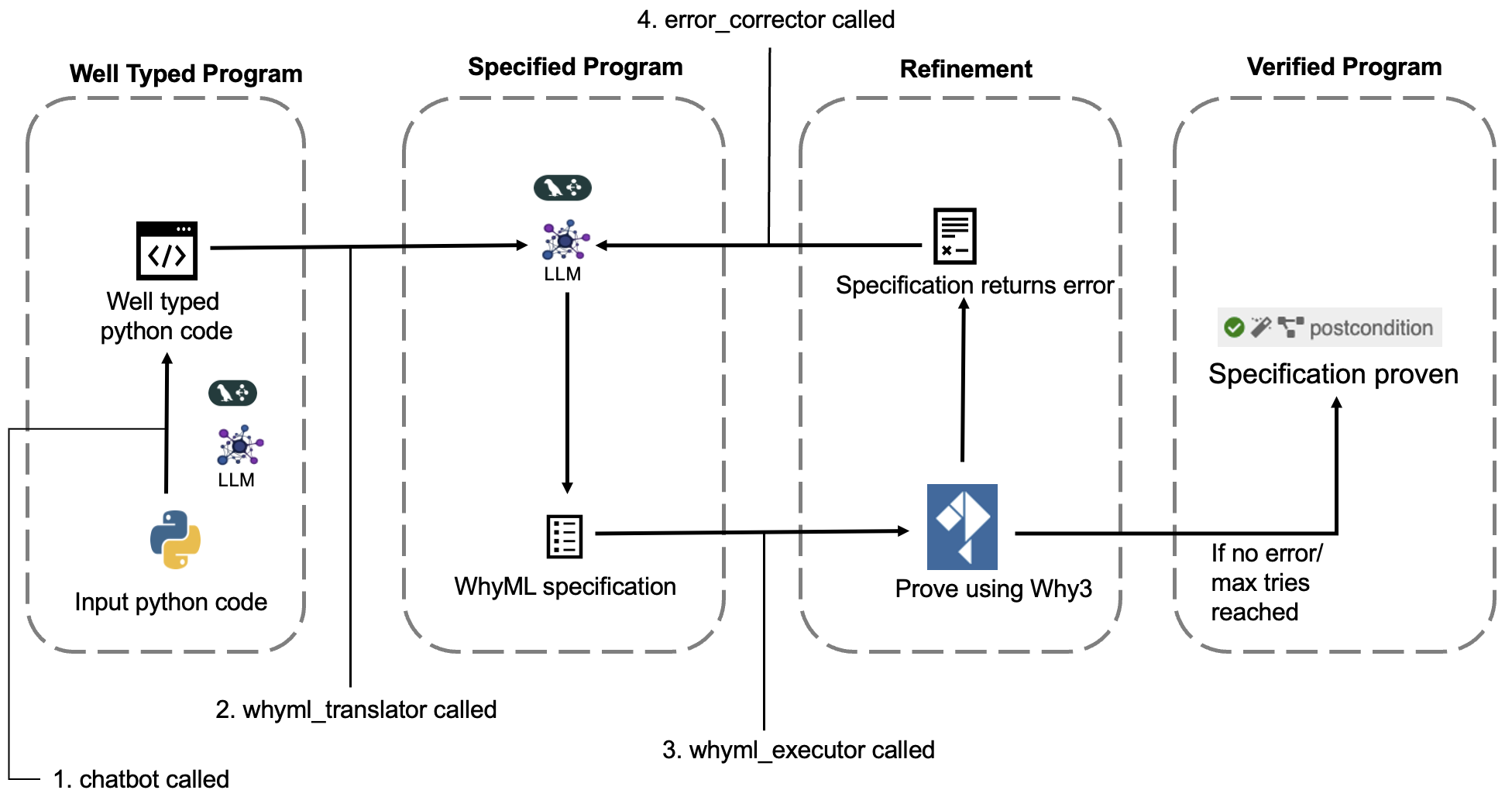}
    \caption{System Architecture (LangGraph) of the Python to WhyML Translation and Verification Framework}
    \label{fig:system_architecture}
\end{figure*}

To empirically investigate the performance of LLMs in the context of formal verification, we designed a closed-loop framework to automate the end to end verification process using LangGraph. Figure~\ref{fig:system_architecture} shows workflow (system architecture) of the framework where an input file (AI-generated Python code) is provided, which is then converted into a well-typed Python program by the LLM. This program then goes through an iterative repair for a maximum of 10 times to generate the final but verifiable WhyML output. At each stage, the LLM generated WhyML specification and error codes are passed back to the LLM along with the well- typed Python code. 

Based on the implementation of a chain of thought process in our methodology, the LLM uses automated reasoning using the \textless thinking\textgreater~block during the error correction, whose nodes are sketched in Table~\ref{tab:langgraph_nodes}. The error\_corrector node is designed to ingest the previous errors and specification, and is explicitly instructed to anlayse the entire history prior to arriving at a solution. 

Since a LangGraph workflow is a stateful directed graph~\cite{b5}, we implemented a graph structure as $G = (V,E)$ where $V$ is the set of vertices and $E$ is the set of directed edges. Each vertex represents a function in the graph.

$V = \{v_{start}, v_{chatbot}, v_{translator}, v_{executor}, v_{corrector}, v_{end}\}$

The directed edges are defined as $E = E_{static} \cup E_{conditional}$, where $E_{static}$ represents the following fixed transitions independent of the state:

\begin{equation*}
\begin{split}
E_{static} = & \{(v_{start}, v_{chatbot}),(v_{chatbot}, v_{translator}), \\
 & (v_{translator}, v_{executor}), (v_{corrector},v_{executor})\}
\end{split}
\end{equation*}

$E_{conditional}$ represents conditional executions depending on the $f(s)$ "should\_retry" function:

\[
  E_{conditional} =
  \begin{cases}
    (v_{executor},v_{corrector}) & \text{ if } f(s) = \text{"retry"}\\
    (v_{executor}, v_{end})  & \text{ if } f(s) = \text{"end"}\\
  \end{cases}
\]

% $E_{conditional} = \{(v_{executor},v_{corrector}) \text{ if } f(s) = \text{"retry"}, 
% (v_{executor}, v_{end}) \text{ if } f(s) = \text{"end"}\}$

\begin{table}[t]
\caption{LangGraph Node Architecture and Functions}
\begin{center}
\tiny
\begin{tabular}{|p{1.4cm}|p{2.8cm}|p{3.6cm}|}
\hline
\textbf{Node} & \textbf{Function} & \textbf{Key Prompt/Logic} \\
\hline
\texttt{chatbot} & LLM: Adds type annotations to Python & "Convert code into well-typed python with proper type hints..." \\
\hline
\texttt{whyml\_translator} & LLM: Translates to WhyML specs & "Expert in Why3 verification. Convert Python to WhyML with contracts..." \\
\hline
\texttt{whyml\_executor} & Why3: Runs prover, captures results & \texttt{why3 prove -P alt-ergo [file.mlw]} \\
\hline
\texttt{error\_corrector} & LLM: Analyses errors, generates fixes based on conditional should\_retry function& "Fix WhyML errors based on message. Analyse previous attempts..." \\
\hline
\end{tabular}
\label{tab:langgraph_nodes}
\end{center}
\end{table}
\begin{figure*}[t]
    \centering
    \includegraphics[scale=0.36]{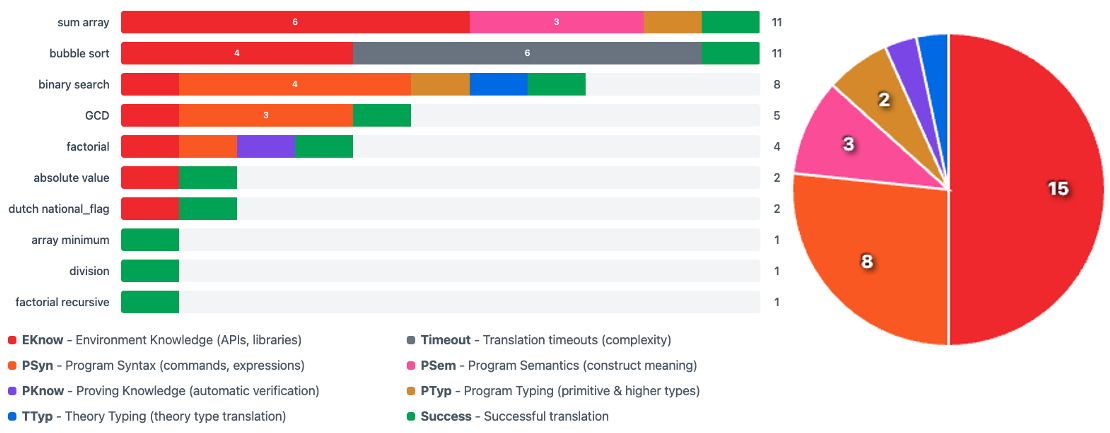}
    \caption{Visualisation of the Capability Gaps}
    \label{fig:visual-gaps}
\end{figure*}
The node is initialised at the START state and then moves to the chatbot node, where the python program undergoes a static translation. The resulting well-typed python code is then passed on to the LLM, which attempts to translate the Python code to a WhyML specified program. The translated code is then sent to the Why3 IDE within a local machine through the whyml\_executor node. This node writes the specified program to a temporary file which uses Why3’s command line interface to verify the specified program using the Alt-Ergo back-end solver. Should there be an error generated by the solver, this node further captures the entire error log. If there is no error, the execution\_success variable is set to True, i.e., no further steps are required and the process moves to the END state.

However, if an error is generated and maximum N (10) tries have not been reached, the conditional edge should\_retry is called. Here, the entire state including the error log, well-typed python code, and translated WhyML specified code is provided back to the LLM keeping the original sequence intact using the error\_corrector node. In this node, a custom prompt is created instructing the LLM to perform an analysis of the errors using the provided history and generate an updated WhyML specified code. The whyml\_executor node is then called on the updated specification node, repeating N times or until the specification is proven. 

\subsection{Capability Gap Analysis}
A capability gap analysis is undertaken on the LLM’s steps to demonstrate the deficiencies in the current LLM knowledge base. As sketched in Table~\ref{tab:capability_gaps}, capabilities are split into multiple dimensions representing the domain (Program, Theory, Specification), the type of error (Typing, Syntax, and Semantics) and the knowledge gap between solvers (Environment, Domain, Proving). As an example, a semantic error will differ between Program, Theory or Specification depending on whether it is LLM-generated in the logical, mathematical, or contractual part of the code.  
\begin{table}[t]
\caption{LLM Capability Gap Classification Framework}
\begin{center}
\tiny
\begin{tabular}{|p{2.8cm}|p{5cm}|}
\hline
\textbf{Code \& Category} & \textbf{Description} \\
\hline
\textbf{PTyp} - Program Typing & If a Python type (e.g., primitive, and higher) is not translated in equivalent Why3 program type \\
\hline
\textbf{TTyp} - Theory Typing & If a Python type is not properly translated in equivalent Why3 theory type \\
\hline
\textbf{STyp} - Specification Typing & If a Python type is not translated in equivalent Why3 specification type \\
\hline
\textbf{PSyn} - Program Syntax & If a syntactic construct of Python program (e.g., command, expressions) is not translated in equivalent Why3 program construct \\
\hline
\textbf{TSyn} - Theory Syntax & If a syntactic construct of Python program is not translated in equivalent Why3 theory construct (e.g., function, lemma, and axioms) \\
\hline
\textbf{SSyn} - Specification Syntax & If a syntactic construct of Python program is not translated in equivalent Why3 specification construct (e.g., pre-, and post condition, variant, invariant, and others) \\
\hline
\textbf{PSem} - Program Semantics & If semantics of a Python program construct is not translated in equivalent Why3 program construct \\
\hline
\textbf{TSem} - Theory Semantics & If semantics of a Python program construct is not translated in equivalent Why3 theory construct \\
\hline
\textbf{SSem} - Specification Semantics & If semantics of a Python program construct is not translated in equivalent Why3 specification construct \\
\hline
\textbf{PKnow} - Proving Knowledge & If proving does not work automatic with the given specification, theory, and program \\
\hline
\textbf{EKnow} - Environment Knowledge & If equivalent translation in Why3 does not properly uses the Why3 environment knowledge (e.g., API, internal and external libraries) \\
\hline
\textbf{DKnow} - Domain Knowledge & If equivalent translation in Why3 does not properly uses domain specific knowledge (e.g., a specific concept knowledge) \\
\hline
\end{tabular}
\label{tab:capability_gaps}
\end{center}
\end{table}
\section{Preliminary Evaluation}

\subsection{Dataset Description}
To determine capability gaps and performance of our implementation, we analysed a total of 10 Python functions. These functions were chosen to cover a variety of common Python language constructs varying in behavioral complexity to determine the reasoning capabilities of the LLM. We split them into 3 categories that determine behavioral complexity to specify and prove its correctness, i.e., basic if-else statements (Easy), to for-loops (Medium) and While-loops (Hard). The performance of the results was assessed by their qualitative and quantitive analysis to get insights on the LLM performance. 
\begin{table*}[htbp]
\caption{Analysis of LLM Errors in Why3}
\begin{center}
\footnotesize
\begin{tabular}{|p{4cm}|p{1.5cm}|p{5.5cm}|p{5.5cm}|}
\hline
\textbf{Error} & \textbf{Capability Gap} & \textbf{Detailed Capability Gap} & \textbf{Function} \\
\hline
unbound function or predicate symbol & EKnow & LLM does not have knowledge of Why3 Libraries and how to use them. & dutch\_national\_flag, absolute\_value, bubble\_sort, sum\_array, binary\_search \\
\hline
Module not found in library & EKnow & LLM does not have knowledge of Why3 Libraries and how to use them. & factorial, bubble\_sort, GCD \\
\hline
Cannot prove termination & PKnow & LLM defined a function but was unable to prove its termination. & factorial \\
\hline
syntax error & PSyn & LLM does not understand WhyML program syntax & GCD \\
\hline
This expression has type int, it cannot be applied & PSyn & LLM does not understand WhyML program syntax & GCD, sum\_array, binary\_search \\
\hline
No suitable match found for notation (=) & PSem & LLM does not know the meaning of the datatypes & sum\_array \\
\hline
This expression has type list... but is expected to have type int & PSem & LLM does not know Type semantics, it is confusing data types & sum\_array, binary\_search \\
\hline
Logical symbol is used in a non-ghost context & EKnow & LLM does not know the difference between program functions and ghost functions. & sum\_array \\
\hline
unbound type symbol & PSem & LLM does not know semantics, it is confusing data types & binary\_search \\
\hline
This term has type int, but is expected to have type real & PTyp & LLM does not know semantics, it is confusing data types & binary\_search \\
\hline
unbound function or predicate symbol '(/)' & PSyn & LLM does not understand why3 program syntax & binary\_search \\
\hline
This expression has type int -> int -> int, but is applied to 3 arguments & PSyn & LLM does not understand why3 program syntax & binary\_search \\
% \hline
% No Errors Found & NA & NA & binary\_search \\
\hline
\end{tabular}
\label{tab:llm_errors}
\end{center}
\end{table*}
\subsection{Quantitative Analysis}
We aggregated the performance of the program to demonstrate the capability gaps between each item. Table~\ref{tab:llm_errors} shows the capability gaps as well as the relevant functions, which are visualised in Figure~\ref{fig:visual-gaps}. The quantitative analysis provided several insights. 

Easy functions such as if-else statements exhibited minimal capability gaps as the LLM provided a suitable WhyML specified program that was accepted by the prover. The number of tries goes up significantly as algorithmic complexity increases. Overall, the LLM was able to provide solutions for 100\% of simple functions with an average rate of 1.5 tries. More complex functions involving for-loops took an average of 4 tries, where the sum\_array function took the full 10 tries to arrive at a correct solution. The hard functions (while-loops with complex invariants/termination terms) took an average of 5.5 tries. This is the first instance where the LLM was unable to provide a sound proof as in the case of bubble\_sort. 

\subsection{Library Knowledge Gaps}
\begin{findingbox}
\textbf{Finding 1:} The most prevalent capability gaps in the LLM are knowledge gaps at circa 50\% of errors.
\end{findingbox}
% \vspace{-0.3cm}
This reveals that the LLM lacks knowledge of Why3's ecosystem. The gap can be addressed using targeted prompting strategies referencing the relevant Why3 library API's. Interestingly, we observed that there was no completely linear relationship between the complexity of the function and number of attempts taken to verify the program. For example, a solution for the division program (easy) was provided in the first attempt, while the absolute\_value (easy) function returned a simple EKnow error in which the LLM did not know to use the real specific operator, and was promptly fixed by the error corrector node. The sum\_array function (medium), consisting of a for-loop took the full 10 attempts to verify the program, while the more complex binary\_search (hard) function which consists of a while-loop took 8 tries. 

\begin{findingbox}
\textbf{Finding 2:} There exists a non-linear relationship between complexity of the python function/program and the number of tries taken by the LLM to generate a automatically verifiable specification.
\end{findingbox}

In the dutch\_national\_flag program, the LLM used incorrect syntax without specifying the relevant library, which suggests that it did not know to use that library. It subsequently increased the complexity by explicitly defining the function but did not import the correct library. This lack of Why3 API knowledge is a recurring theme in programs of any complexity. 

Environmental knowledge gaps were also prevalent in the factorial function, where the LLM referenced a non-existent library within the system. Then it tried to fix this by removing the reference and adding a custom function that included recursion. The LLM was unable to prove termination of the recursive function and reverted to a simpler loop structure. 

In bubble\_sort the LLM again referenced a non-existent library, manually defined a function and created a highly complex function. The prover timed out after the max (10) tries since the LLM was unable to verify the program. 

\begin{findingbox}
\textbf{Finding 3:} LLM is unable to directly translate or prove highly complex postconditions, and reverts to the simplest specification option. 
\end{findingbox}

\subsection{Syntactic Gaps}
The results show that roughly 30\% of the gaps occurring in the programs were caused by misunderstanding the syntax of WhyML by the LLM. In the GCD function, the LLM initially used incorrect syntax for the modulo operator. It then altered to function using an incorrect division module without correctly identifying the underlying problem. The LLM repeated this error one more time before it corrected itself using the error\_corrector loop to provide a correct solution. These minoro errors, took several attempts for the LLM, suggesting a lack of syntactic understanding and a lack of context fed back by the Why3 error logs.

\begin{findingbox}
\textbf{Finding 4:} The context provided by the WhyML error logs is not optimised for LLM performance in a self-correcting loop. 
\end{findingbox}

In binary\_search, the LLM confused the "/" python division operator and placed it in the Why3 syntax. It then incorrectly defined a function as "int,int,int" with two input parameters, but provided three arguments. In this case, the lambda notation was not understood and was corrected by changing the function call to match the signature. This demonstrates that the LLM has limited knowledge of WhyML's lambda expression syntax.

\subsection{Semantic and Typing Gaps}
Semantic and Type errors accounted for 19\% and 3\% respectively, indicating that LLMs are not particularly prone to making these types of errors. This could also be explained by the prevalence of knowledge gaps with the LLM. Should improvements be made to the LLMs understanding of the Why3 Libraries in the form of additional context or fine tuning, occurrences of these errors may increase in future iterations. 

In sum\_array, the LLM was initially confused between the list and array semantics of data-types. It lacked an understanding of the meaning of the different data types. This occurred since the initial Python program consisted of the keyword "list" as a variable name, which the NLP repeated without understanding the semantics, causing the LLM to keep nesting list types. This followed a type error (PTyp), in which the LLM created a list type but expected an int type, indicating an incorrect type translation from the python program. The final successful attempt abandoned the iterative approach and defined sum\_array as a simple recursive function. This suggests that the LLM is unable to handle the complexity of proving multiple postconditions.

In binary\_search, type errors occurred at the first attempt, where Python's float type was not translated into the equivalent theory type in Why3. The LLM tried rectifying this by adding real numbers, which also did not produce a valid specification. It then finally added integers which was successfully proven by the SMT solver. This suggests that the LLM is not very strong in translating between types.

\end{document}